\documentclass[11pt,twoside]{article}
\usepackage{asp2006}
\usepackage{psfig}
\usepackage{epsf}
\usepackage{graphics}
\usepackage{lscape}
\def\edcomment#1{\iffalse\marginpar{\raggedright\sl#1\/}\else\relax\fi}
\reversemarginpar

\begin{document}
\title{Quasar Metal Abundance and FIR Luminosity}
\author{L. E. Simon and F. Hamann}
\affil{Department of Astronomy, University of Florida, 211 Bryant Space Scinece Center, PO Box 112055, Gainesville FL, 32611, USA}

\begin{abstract}
  We compare the metallicities in high-redshift quasars to the star
  formation rates (SFR) in their host galaxies using measurements of
  broad emission lines and far-infrared (FIR) luminosities.  The FIR
  emission indicates the level of ongoing massive starbursts in the
  galaxy, whereas the abundance of metals in the gas surrounding the
  quasar indicates the amount of star formation which occurred before
  the visible quasar phase began.  The results of this study can be
  used to constrain the late stages of starburst-quasar evolution. We
  detect high metallicities throughout the sample, up to several times
  solar, confirming that star formation must have begun before the
  visible quasar phase.  However, we do not detect a trend in
  metallicity versus current SFR.
\end{abstract}

\vspace{-0.5cm}
\section{Introduction}

High redshift quasars are thought to represent an early stage of
galaxy evolution, where major mergers trigger violent star formation
and the rapid growth of a central super-massive black hole.  However,
the timing of the quasar phase during a galaxy's evolution is not well
understood.  For example, the quasar could trigger a massive burst of
star formation in the host galaxy, quench star formation in the host,
or have no direct effect on star formation whatsoever.

The gas-phase abundance of the near-quasar environment has been probed
out to high redshifts, particularly through the use of quasar broad
emission lines, and has consistently shown metal abundances near or
above the solar value.  This result suggests that there is always
significant star formation enriching the surrounding gas before the
black hole becomes a luminous quasar.

One possible indicator of ongoing star formation is emission at
far-infrared (FIR) wavelengths, where FIR emission is caused by dust
heated by ultraviolet (UV) emission from massive starbursts (Kennicutt
1998, and references therein). A recent analysis by Hao et al. (2008)
of FIR data from the literature indicates that high-redshift quasars
can have star formation rates (SFR) up to
~$>$1000~M$_{\odot}$~yr$^{-1}$~ for the brightest L$_{FIR}$~ quasars.

We look for signs of age progression in various stages of the quasar
life-cycle by measuring the metallicity of the quasar emission line
gas at different increments of L$_{FIR}$, i.e., ongoing star formation.
Presumably the later stages of evolution are older and therefore show
more metal enrichement than earlier, younger stages.  To smooth over
variations in individual quasars, we make composite spectra of quasars
in each of three L$_{FIR}$ increments.

We measure metallicites using the N~V $\lambda$1240 and C~IV
$\lambda$1549 UV emission line strengths.  The ratio of N~V to C~IV
increases with metallicity (and age), due to the secondary enrichment
of nitrogen (Hamann et al. 2002).  Thus, a trend in the N~V/C~IV ratio
over the three L$_{FIR}$ increments could indicate a progression from
one evolutionary phase to the next.

\section{Method}

We compile a sample of 30 optical spectra from SDSS of known optically
bright FIR-observed quasars with a redshift range of 2~$<$~z~$<$~5.
This redshift range shifts the UV region of the quasar spectrum into
the optical region observed by SDSS.  We use the L$_{FIR}$ from Hao et
al. (2008), who compile a list of optically selected quasars observed
at FIR wavelengths, and we use the optical (B) magnitudes from Omont
et al. (2001, 2003) and Carilli et al. (2001) who actually observed
these sources in the FIR using MAMBO and SCUBA respectively.  We sort
the SDSS spectra by L$_{FIR}$ into three non-overlapping categories of
$\sim$10 spectra each with the brightest category containing spectra
with ~$log(L_{60\mu m}/L_{\odot})~>$~13.2, the faint category
containing spectra with ~12.8~$<~log(L_{60\mu m}/L_{\odot})~<$~13.15
and the category not detected in the FIR with ~$log(L_{60\mu
  m}/L_{\odot})~<$~12.8 upper limits.  The range of M$_B$ is -26.5 to
-28.9.  All three categories span this range, but the FIR bright
category is dominated by M$_B~>$~-28, whereas the other two categories
are clustered near M$_B$~-27.4.

We create average combined spectra of each sample, removing strong
absorption by hand before combining.  We normalize the combined
spectra using power law fits to regions relatively free of emission or
absorption; rest-frame 1447-1473~\AA, 1757-1783~\AA~ and 1987-2013~\AA~
for the non-detected in FIR composite.  We intend to use gaussian
profiles to fit N~V, C~IV and Ly$\alpha$ emission in each composite
spectrum.

\section{Results}

Based on preliminary analysis, the N~V/C~IV line ratios in all three
composites indicate that the quasar environments are metal rich, with
approximately a few times solar metallicities. A weak trend also
appears for increasing metallicity (i.e., increasing N~V/C~IV line
ratio) in the sources with larger L$_{FIR}$ (Figure~1). However, these
sources also tend to have larger accretion luminosities, L$_{UV}$, and
their slightly higher metallicities can be attributed to a known
correlation between metallicity, L$_{UV}$ and central black hole mass
among quasars generally (Warner, Hamann, \& Dietrich, 2004). Thus, we
find no clear evidence for evolutionary (enrichment) differences
between quasars sorted by the presumed SFR indicator,
L$_{FIR}$. Instead, we find simply an affirmation of previous work
showing that quasar broad emission line regions are generically metal
rich, with the highest metallicities occuring in the most luminous and
most massive quasar environments.

These results can be understood in the context of galaxy evolution and
black hole growth by noting that the SMBHs that power these quasars
have masses of order $10^9~M_{\odot}$ or higher. The metal rich gas
suggests vigorous previous star formation, as is expected along the
way toward making these massive SMBHs.  The lack of an obvious trend
in line strengths with L$_{FIR}$ may indicate that ongoing star
formation in quasar hosts is not a significant source of enrichment
for the gas. This result is still consistent with current models where
quasars are preceded by massive starbursts in their hosts (Hopkins et
al. 2008).

\begin{figure}
\plotone{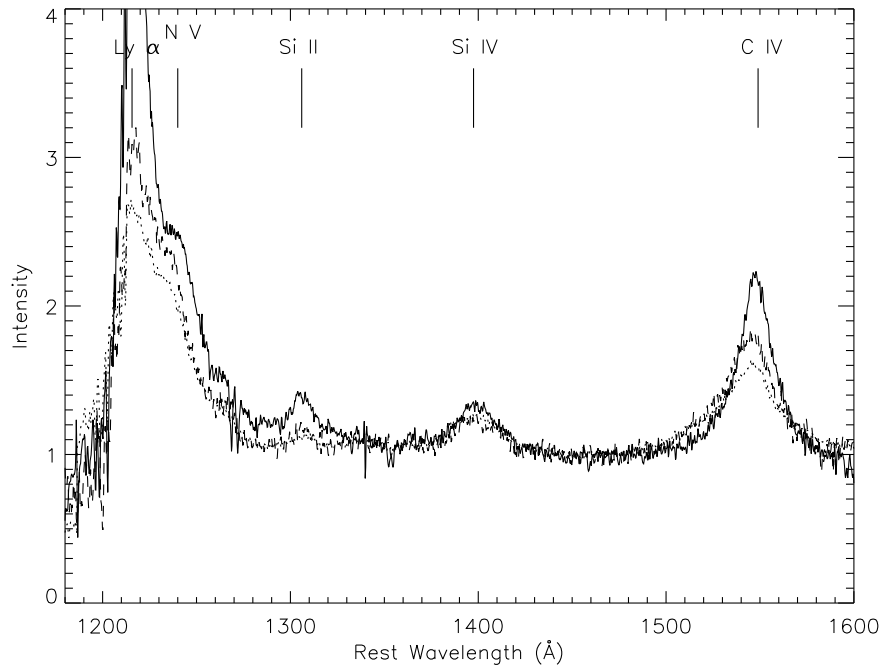}
\caption{Composite SDSS quasar spectra grouped by $L_{FIR}$, shifted
  to rest-UV wavelengths.  The solid curve represents the composite of
  upper-limit non-detections in FIR, the dashed curve represents the
  composite of faint FIR quasars and the dotted curve represents the
  composite of bright FIR quasars. Promonent emission lines are
  labeled. All composite spectra are normalized using a power law fit
  to feature-free regions as described in the text. }
\label{fig:fits}
\end{figure}

\acknowledgements This work was supported in part by NASA/Chandra
award G08-9102X.

\end{document}